\documentclass[preprint,showpacs,preprintnumbers,amsmath,amssymb]{revtex4}

\usepackage{graphicx}% Include figure files
\usepackage{dcolumn}% Align table columns on decimal point
\usepackage{bm}% bold math

\begin{document}

\title{Second post-Newtonian approximation of Einstein-aether theory}

\author{Yi Xie$^1$}
\email{yixie@nju.edu.cn}
\author{Tian-Yi Huang$^{1,2}$}
\email{tyhuang@nju.edu.cn}
\affiliation{
$^1$Department of Astronomy, Nanjing University, Nanjing 210093, China\\
$^2$Shanghai Astronomical Observatory, Chinese Academy of Sciences, Shanghai 20030, China
}

\date{\today}

\begin{abstract}
	In this paper, second post-Newtonian approximation of Einstein-aether theory is obtained by Chandrasekhar's approach. Five parameterized post-Newtonian parameters in first post-Newtonian approximation are presented after a time transformation and they are identical with previous works, in which $\gamma=1$, $\beta=1$ and two preferred-frame parameters remain. Meanwhile, in second post-Newtonian approximation, a parameter, which represents third order nonlinearity for gravity, is zero the same as in general relativity. For an application for future deep space laser ranging missions, we reduce the metric coefficients for light propagation in a case of $N$ point masses as a simplified model of the solar system. The resulting light deflection angle in second post-Newtonian approximation poses another constraint on the Einstein-aether theory.
\end{abstract}

\pacs{04.50.+h, 04.25.Nx, 04.80.Cc}
\maketitle

\allowdisplaybreaks

\section{Introduction}

Although Einstein's general relativity (GR) has achieved great success both in experimental tests and in astrophysical applications during the last few decades, the desire to find a gravitation theory consistent with quantum theory together with the ever-increasing precision of experiments and astrometric observations has urged many ``alternative theories'' to be proposed. Among them, vector-tensor theories are usually investigated in the research of preferred frames and violations of Lorentz invariance, due to the existence of a vector field $K_{\mu}$ in gravity besides a metric tensor $g_{\mu\nu}$. 

Vector fields without a constraint were considered by Will and Nordtvedt for preferred-frame theories of gravity in the 1970s \cite{Will1972,Nordtvedt1972}. They introduced three parameters $\alpha_1$, $\alpha_2$ and $\alpha_3$ to describe preferred-frame effects. But, all previous preferred-frame theories were ruled out by gravimeter data. After that, Hellings and Nordtvedt investigated a massless vector field in addition to the metric field in the solar system experiments and cosmological expansion \cite{Hellings1973}. In \cite{TEGP}, a summary of previous works is given and a general action of a vector-tensor theory without a constraint is proposed. However, it is shown that an unconstrained condition on the vector norm induces instabilities in those theories \cite{Elliott2005}.

On the other hand, vector-tensor theories with a potential leading to violations of Lorentz invariance were studied by several authors \cite{Kostelecky1989, Clayton1999, Clayton2000}. A background dynamical tensor field can also break Lorentz symmetry. The simplest cases are scalar fields with a non-zero gradient and vector fields. A special case is a unit timelike vector field, which is called ``Einstein-aether theory'' or ``ae-theory'' for short \cite{Jacobson2001}. This theory has been intensively investigated by several authors in the past few years (see \cite{Eling2006review} for a review). The linearized theory of ae-theory and the propagation of aether waves are studied in \cite{Jacobson2004}. In the aspect of post-Newtonian (PN) approximation (a weak-field and slow-motion limit), two Eddington-Robertson-Schiff parameters, $\gamma$ and $\beta$, are obtained by applying the asymptotic weak field limit of spherically symmetric static solutions \cite{Eling2004}. One of the parameters due to the preferred frame, $\alpha_2$, is calculated in the low-energy effective theory \cite{Graesser2005}. Then, the parameterized post-Newtonian (PPN) formalism is used to attain 10 PPN parameters and other constraints are found \cite{Foster2006a}. Also, the radiation damping in ae-theory is investigated in \cite{Foster2006b,Foster2006bE}. In astrophysics application, although it is shown that there is nonexistence of pure aether stars, regular perfect fluid star solutions exist with static aether exteriors \cite{Eling2006a}. In addition, black holes are studied in ae-theory and they are found to be very close to Schwarzschild solution outside its horizon and have a spacelike singularity inside \cite{Eling2006b}. Furthermore, numerical simulations of gravitational collapse in ae-thoery are performed, in which stationary black holes would appear as long as the aether coupling constants are not too large \cite{Garfinkle2007}. For the properties of non-rotating neutron stars in ae-theory, it is shown that it leads to lower maximum neutron star masses, as well as larger surface redshifts at a particular mass, for a given nuclear equation of state \cite{Eling2007}. Strong field effects on binary system are also considered in ae-theory. There exists a one-parameter family with ``small-enough'' couplings, which passes all current observational tests \cite{Foster2007}.

To test vector-tensor theories in the solar system by future high-precision experiments, the PN approximation of these theories is needed. Therefore PN corrections of equations of motion, equations of light, and other relativistic effects should be derived for experiments. Two approaches can achieve this task. One is the PPN formalism in first post-Newtonian (1PN) approximation proposed by Will and Nordtvedt \cite{Will1972, TEGP}. In this framework, 1PN metric is parameterized with 10 PPN parameters, and the differences among different theories of gravity are represented by the values of these parameters. In \cite{Will1972,Nordtvedt1972, Hellings1973, Foster2006a}, the PPN approach is used to attain the 1PN approximation of vector-tensor theories in unconstrained and constrained cases. Although there are some efforts to extend this formalism to second post-Newtonian (2PN) approximation by Nordtvedt and Benacquista in \cite{Benacquista1988,Benacquista1992,Nordtvedt1993}, which introduce a lot of parameters to cover various relativistic theories, the ability of such an approach to describe the physical features at the 2PN level is unclear (see a brief comment in \cite{Damour1996}). In this work, we focus only on ae-theory, therefore we employ a ``theory-dependent'' approach. This approach, which solves the field equations through iteration, is proposed by Chandrasekhar \cite{Chandrasekhar1965,Chandrasekhar1969}, who obtained 1PN and 2PN approximation of GR.

In what follows, our conventions and notations generally follow those of \cite{mtw}. The signature of metric is $(-,+,+,+)$. Greek indices take the values from $0$ to $3$, Latin indices take the values from $1$ to $3$ and repeated indices mean Einstein's summation. Bold letters $\bm{A}=(A^i) $ denote spatial vectors. A dot between two spatial vectors, $\bm{A}\cdot\bm{B}$, means the Euclidean scalar product.

\section{Action and field equations}
\label{actionfe}
In a general tensor-vector theory of gravity, the Lagrangian scalar density involves a metric $g_{\mu \nu}$ and a 4-vector field $K_{\mu}$. The action defining the theory reads
\begin{eqnarray}
	\label{action}
	S&=&\frac{c^3}{16\pi G}\int \bigg[\phantom{+}f_0(K^2)R+ f_1(K^2)K^{\mu;\nu}K_{\mu;\nu}+f_2(K^2)K^{\mu}_{;\mu}K^{\nu}_{;\nu}+f_3(K^2)K^{\mu;\nu}K_{\nu;\mu}\nonumber\\
	&&\phantom{\frac{c^3}{16\pi G}\int \bigg[}+f_4(K^2)K^{\lambda}K^{\kappa}_{;\lambda}K^{\rho}K_{\kappa;\rho}+\lambda\bigg(\frac{K^2}{\phi^2}+1\bigg)\bigg]\sqrt{-g}\mathrm{d}^4x+S_m(\psi,g_{\mu \nu}),
\end{eqnarray}
where $c$ is the ultimate speed of the special theory of relativity, $G$ is an a priori gravitational constant, $g=\mathrm{det}(g_{\mu \nu})<0$ is the determinant of the metric tensor $g_{\mu \nu}$, $R$ is the Ricci scalar, $\psi$ denotes all the matter fields, $K^2\equiv K^{\lambda}K_{\lambda}$ and $-\phi\delta^0_{\mu}$, where $\delta^0_{\mu}$ is Kronecker $\delta$, is the asymptotic value of $K_{\mu}$. The Lagrange multiplier $\lambda$ constrains the vector field $K^2$ to be $-\phi^2$. Here, we respect the Einstein equivalence principle so that the matter fields $\psi$ do not interact with the vector field, i.e. the action of matter $S_m$ is the function of $\psi$ and $g_{\mu\nu}$ only.

Variations of $g_{\mu \nu}$ and $K_{\mu}$ give the field equations
\begin{equation}
	\label{tfe}
	\Theta^{(0)}_{\mu \nu}+\Theta^{(1)}_{\mu \nu}+\Theta^{(2)}_{\mu \nu}+\Theta^{(3)}_{\mu \nu}+\Theta^{(4)}_{\mu \nu}+\Theta^{(5)}_{\mu\nu}=\frac{8\pi G}{c^2}T_{\mu \nu},
\end{equation}
and
\begin{equation}
	\label{vfe}
	\Xi^{\mu}_{(0)}+\Xi^{\mu}_{(1)}+\Xi^{\mu}_{(2)}+\Xi^{\mu}_{(3)}+\Xi^{\mu}_{(4)}+\Xi^{\mu}_{(5)}=0,
\end{equation}
where
\begin{eqnarray}
	\label{}
	\Theta^{(0)}_{\mu \nu} & = & f_0R_{\mu \nu}-\frac{1}{2}f_0g_{\mu \nu}R\nonumber\\
	&&+g_{\mu \nu} \Box_g f_0-(f_0)_{;\mu\nu}+f'_0RK_{\mu}K_{\nu},\\
	\label{}
	\Theta^{(1)}_{\mu\nu}&=&f_1K_{\mu;\lambda}K_{\nu}^{\phantom{\nu};\lambda}+f_1K_{\lambda;\mu}K^{\lambda}_{\phantom{\lambda};\nu}\nonumber\\
	&&-\frac{1}{2}f_1g_{\mu \nu}K^{\lambda;\rho}K_{\lambda;\rho}+f'_1K_{\mu}K_{\nu}K^{\lambda;\rho}K_{\lambda;\rho}\nonumber\\
	&&+[f_1K^{\lambda}K_{(\mu;\nu)}]_{;\lambda}-[f_1K^{\lambda}_{\phantom{\lambda};(\mu}K_{\nu)}]_{;\lambda}-[f_1K_{(\mu}K_{\nu)}^{\phantom{\nu)};\lambda}]_{;\lambda},\\
	\label{}
	\Theta^{(2)}_{\mu \nu} & = & 2f_2K^{\lambda}_{\phantom{\lambda};\lambda}K_{(\mu;\nu)}-2(f_2K^{\lambda}_{\phantom{\lambda};\lambda}K_{(\mu})_{;\nu)}+g_{\mu\nu}(f_2K^{\lambda}_{\phantom{\lambda};\lambda}K_{\rho})^{;\rho}\nonumber\\
	&&-\frac{1}{2}f_2g_{\mu \nu}K^{\lambda}_{\phantom{\lambda};\lambda}K^{\rho}_{\phantom{\rho};\rho}+f'_2K_{\mu}K_{\nu}K^{\lambda}_{\phantom{\lambda};\lambda}K^{\rho}_{\phantom{\rho};\rho},\\
	\label{}
	\Theta^{(3)}_{\mu\nu} & = & 2f_3K^{\lambda}_{\phantom{\lambda};(\mu}K_{\nu);\lambda}-\frac{1}{2}f_3g_{\mu\nu}K^{\lambda;\rho}K_{\rho;\lambda}+f_3'K_{\mu}K_{\nu}K^{\lambda;\rho}K_{\rho;\lambda}\nonumber\\
	&&+[f_3K^{\lambda}K_{(\mu;\nu)}]_{;\lambda}-[f_3K^{\lambda}_{\phantom{\lambda};(\mu}K_{\nu)}]_{;\lambda}-[f_3K_{(\mu}K_{\nu)}^{\phantom{\nu)};\lambda}]_{;\lambda},\\
	\Theta^{(4)}_{\mu \nu} & = & 2f_4K^{\lambda}K^{\rho}_{\phantom{\rho};\lambda}K_{\rho;(\mu}K_{\nu)}+f_4K^{\lambda}K^{\rho}K_{\mu;\lambda}K_{\nu;\rho}\nonumber\\
	&&-\frac{1}{2}f_4g_{\mu\nu}K^{\lambda}K^{\rho}K^{\kappa}_{\phantom{\kappa};\lambda}K_{\kappa;\rho}-(f_4K_{\mu}K_{\nu}K^{\rho}K^{\lambda}_{\phantom{\lambda};\rho})_{;\lambda}\nonumber\\
	&&+f'_4K_{\mu}K_{\nu}K^{\lambda}K^{\rho}K^{\kappa}_{\phantom{\kappa};\lambda}K_{\kappa;\rho},\\
	\Theta^{(5)}_{\mu\nu} & = & \frac{\lambda}{2\phi^2}[2K_{\mu}K_{\nu}-g_{\mu\nu}(K^2+\phi^2)],\\
	\Xi^{\mu}_{(0)} & = & f'_0K^{\mu}R,\\
	\Xi^{\mu}_{(1)} & = & f'_1K^{\mu}K^{\lambda;\rho}K_{\lambda;\rho}-(f_1K^{\mu;\lambda})_{;\lambda},\\
	\Xi^{\mu}_{(2)} & = & f'_2K^{\mu}K^{\lambda}_{\phantom{\lambda};\lambda}K^{\rho}_{\phantom{\rho};\rho}-(f_2K^{\lambda}_{\phantom{\lambda};\lambda})^{;\mu},\\
	\Xi^{\mu}_{(3)} & = & f'_3K^{\mu}K^{\lambda;\rho}K_{\rho;\lambda}-(f_3K^{\lambda;\mu})_{;\lambda},\\
	\Xi^{\mu}_{(4)} & = & f'_4K^{\mu}K^{\lambda}K^{\rho}K^{\kappa}_{\phantom{\kappa};\lambda}K_{\kappa;\rho}+f_4K^{\rho}K^{\lambda;\mu}K_{\lambda;\rho}\nonumber\\
	&&-(f_4K^{\lambda}K^{\rho}K^{\mu}_{\phantom{\mu};\rho})_{;\lambda},\\
	\label{eqlambda1}\Xi^{\mu}_{(5)} & = & \frac{\lambda}{\phi^2}K^{\mu},
\end{eqnarray}
in which $\Box_g(\cdot)\equiv (\cdot)_{;\mu\nu}g^{\mu\nu}$, $f'_{\mu}\equiv\partial f_{\mu}/\partial (K^2)\,(\mu=0,1,2,3,4)$ and parentheses surrounding a group of indices mean symmetrization, for example, $K_{(\mu;\nu)}=(1/2)(K_{\mu;\nu}+K_{\nu;\mu})$. The energy-momentum tensor $T^{\mu\nu}$ is
\begin{equation}
	\label{}
	c^2T^{\mu\nu}\equiv-\frac{2c}{\sqrt{-g}}\frac{\partial S_m(\psi,g_{\mu\nu})}{\partial g_{\mu\nu}},
\end{equation}
and $T_{\mu\nu}=g_{\mu\sigma}g_{\nu\rho}T^{\sigma\rho}$. Following \cite{BD1989,Damour1992,Damour1996}, we define mass, current and stress density as
\begin{eqnarray}
	\label{defsigma}
	\sigma & \equiv & T^{00}+T^{kk},\\
	\label{defsigmai}
	\sigma_i & \equiv & cT^{0i},\\
	\label{defsigmaij}
	\sigma_{ij} & \equiv & c^2T^{ij}.
\end{eqnarray}
Another way to define $\sigma$ involving the PPN parameters $\gamma$ and $\beta$ in 1PN is \cite{Klioner2000}
\begin{equation}
	\label{}
	\sigma=T^{00}+\gamma T^{kk}+\frac{1}{c^2}T^{00}(3\gamma-2\beta-1)U+\mathcal{O}(c^{-4}),
\end{equation}
where $U$ is the Newtonian potential. Due to $\gamma=1$ and $\beta=1$ in ae-theory (see a PPN parameter summary in Sec. \ref{sumppn}), these two definitions are equivalent to each other in 1PN approximation. It is worth emphasizing that, in these definitions, the matter is described by the energy-momentum tensor without specific equation of state.

Contracting the field equation (\ref{tfe}) and substituting it into the field equations (\ref{tfe}) and (\ref{vfe}), it leads to
\begin{eqnarray}
	\label{Ruv}
	R_{\mu \nu} & = & \frac{8\pi G}{c^2}\theta(K^2)(T_{\mu\nu}+f_{\mu\nu}T)\nonumber\\
	&&-\theta(K^2)f_{\mu\nu}\bigg(3\Box_gf_0+\sum_{i=1}^{5}\Theta^{(i)}\bigg)\nonumber\\
	&&+\theta(K^2)\bigg[(f_0)_{;\mu\nu}-g_{\mu\nu}\Box_gf_0-\sum_{i=1}^{5}\Theta^{(i)}_{\mu\nu}\bigg],
\end{eqnarray}
and
\begin{equation}
	\label{}
	\bigg[3\Box_gf_0+\sum_{i=1}^{5}\Theta^{(i)}\bigg]\eta(K^2)K^{\mu}+\sum_{i=1}^{5}\Xi^{\mu}_{(i)}=\frac{8\pi G}{c^2}\eta(K^2)TK^{\mu},
\end{equation}
where $\Theta^{(i)}\equiv\Theta^{(i)}_{\mu\nu}g^{\mu\nu}$, $T\equiv T_{\mu\nu}g^{\mu\nu}$, and the coupling functions are
\begin{eqnarray}
	\label{}
	\theta(K^2) & \equiv & \frac{1}{f_0},\\
	\eta(K^2) & \equiv & \frac{f'_0}{f_0-f'_0K^2},
\end{eqnarray}
and 
\begin{equation}
	\label{}
	f_{\mu\nu}\equiv\frac{1}{f_0-f'_0K^2}\bigg(f'_0K_{\mu}K_{\nu}-\frac{1}{2}f_0g_{\mu\nu}\bigg).
\end{equation}

According to previous works, we assume the action parameters are
\begin{eqnarray}
	\label{}
	f_0 & = & 1-c_0\frac{K^2}{\phi^2},\\
	f_1 & = & -\frac{c_1}{\phi^2},\\
	f_2 & = & -\frac{c_2}{\phi^2},\\
	f_3 & = & -\frac{c_3}{\phi^2},\\
	f_4 & = & \frac{c_4}{\phi^2},
\end{eqnarray}
where $c_{\mu}\,(\mu=0,1,2,3,4)$ are constants.  When $c_4=0$ and $\lambda=0$, the action (\ref{action}) reduces to the case in \cite{Will1972,Hellings1973,TEGP}; When $c_0=0$, $\phi=1$ and $\lambda$ plays a role of the Lagrange multiplier to constrain $K^2=-1$, the action (\ref{action}) becomes the ae-theory \cite{Jacobson2001,Jacobson2004,Eling2004,Foster2005,Eling2006a,Eling2006b,Foster2006a,Foster2006b} (see Table \ref{tab1}). 
\begin{table}
	\caption{\label{tab1} Special cases of the action (\ref{action})}
\begin{ruledtabular}
\begin{tabular}{ll}
Special case &parameters\\
\hline
General Relativity & $c_0=c_1=c_2=c_3=c_4=0$\\
Einstein-Maxwell theory & $c_1+c_3=0$, $c_0=c_2=c_4=0$\\
Will \& Nordtvedt \cite{Will1972} & $c_0=c_2=c_3=c_4=0$\\
Hellings \& Nordtvedt \cite{Hellings1973} & $c_1+c_2+c_3=0$, $c_4=0$\\
Will \cite{TEGP} & $c_4=0$\footnote{$c_0=-\omega$, $c_1=2\epsilon-\tau$, $c_2=-\eta$, $c_3=\eta-2\epsilon$ \cite{Eling2004}.}\\
Aether theory \cite{Jacobson2001,Jacobson2004,Eling2004,Foster2005,Eling2006a,Eling2006b,Foster2006a, Foster2006b}& $c_0=0$, $K^2=-1$\\
\end{tabular}
\end{ruledtabular}
\end{table}

In this paper, we concentrate on ae-theory only, that is, $f_0=1$, $f_1=-c_1$, $f_2=-c_2$, $f_3=-c_3$ and $f_4=c_4$. Corresponding field equations of ae-theory are simplified as
\begin{equation}
	\label{aetfe}
	R_{\mu \nu}  =  \frac{8\pi G}{c^2}\bigg(T_{\mu\nu}-\frac{1}{2}g_{\mu\nu}T\bigg)+\frac{1}{2}g_{\mu\nu}\sum_{i=1}^{5}\Theta^{(i)}-\sum_{i=1}^{5}\Theta^{(i)}_{\mu\nu},
\end{equation}
and
\begin{equation}
	\label{aevfe}
	\sum_{i=1}^{5}\Xi^{\mu}_{(i)}=0,
\end{equation}
and the Lagrange multiplier $\lambda$ from Eqs. (\ref{eqlambda1}) and (\ref{aevfe}) can be expressed as
\begin{equation}
	\label{lambdaexp}
	\lambda=\sum_{i=1}^4\Xi^{\alpha}_{(i)}K_{\alpha},
\end{equation}
where $K^2=-1$ is used. These field equations coincide with previous works' \cite{Jacobson2001,Jacobson2004,Eling2004,Foster2005,Eling2006a,Eling2006b,Foster2006a,Foster2006b}.

\section{PN expansion of the Metric and Vector field}
\label{metric}
In PN approximation, we consider an asymptotically flat spacetime, whose metric $g_{\mu\nu}$ to second order has the form as
\begin{eqnarray}
	\label{PNmetric1}
	g_{00} & = & -1+\epsilon^2\overset{(2)}{h}_{00}+\epsilon^4\overset{(4)}{h}_{00}+\epsilon^6\overset{(6)}{h}_{00}+\mathcal{O}(\epsilon^8),\\
	\label{PNmetric2}
	g_{0i} & = & \epsilon^3\overset{(3)}{h}_{0i}+\epsilon^5\overset{(5)}{h}_{0i}+\mathcal{O}(\epsilon^7),\\
	\label{PNmetric3}
	g_{ij} & = & \delta_{ij}+\epsilon^2\overset{(2)}{h}_{ij}+\epsilon^4\overset{(4)}{h}_{ij}+\mathcal{O}(\epsilon^6),
\end{eqnarray}
where $\epsilon\equiv1/c$. Furthermore, we simplify the notations with the definitions \cite{Kopeikin2004}:
\begin{equation}
	N\equiv\overset{(2)}{h}_{00}, \qquad L\equiv\overset{(4)}{h}_{00}, \qquad L_i\equiv\overset{(3)}{h}_{0i}, \qquad H_{ij} \equiv\overset{(2)}{h}_{ij}, \qquad H\equiv\overset{(2)}{h}_{kk},
\end{equation}
\begin{equation}
	Q\equiv\overset{(6)}{h}_{00},  \qquad Q_i\equiv\overset{(5)}{h}_{0i}, \qquad Q_{ij} \equiv\overset{(4)}{h}_{ij}.
\end{equation}
And the expansions of vector field are
\begin{eqnarray}
	\label{}
	K_0 & = & -1+\epsilon^2\overset{(2)}{K}_0+\epsilon^4\overset{(4)}{K}_0+\epsilon^6\overset{(6)}{K}_0+\mathcal{O}(\epsilon^8),\\
	K_i & = & \epsilon^3\overset{(3)}{K}_i+\epsilon^5\overset{(5)}{K}_i+\mathcal{O}(\epsilon^7),
\end{eqnarray}
which is a timelike unit vector.
Hence,
\begin{eqnarray}
	\label{K2}
	K^2 & = & -1-\epsilon^2\bigg(N-2\overset{(2)}{K}_0\bigg)-\epsilon^4\bigg(N^2+L-2N\overset{(2)}{K}_0+{\overset{(2)}{K}_0}^2-2\overset{(4)}{K}_0\bigg)\nonumber\\
	&&-\epsilon^6\bigg[N^3+2NL-L_kL_k+Q+N{\overset{(2)}{K}_0}^2-2N\overset{(4)}{K}_0-2(N^2+L)\overset{(2)}{K}_0\nonumber\\
	&&\phantom{+\epsilon^6\bigg[}+2L_k\overset{(3)}{K}_k-\overset{(3)}{K}_k\overset{(3)}{K}_k+2\overset{(2)}{K}_0\overset{(4)}{K}_0-2\overset{(6)}{K}_0\bigg].
\end{eqnarray}
Consequently, with the unitary and timelike condition $K^2=-1$, $\overset{(2)}{K}_0$, $\overset{(4)}{K}_0$ and $\overset{(6)}{K}_0$ can be solved at corresponding orders as
\begin{equation}
	\label{Kt2}
	\overset{(2)}{K}_0=\frac{N}{2},
\end{equation}
\begin{equation}
	\label{}
	\overset{(4)}{K}_0=\frac{L}{2}+\frac{N^2}{8},
\end{equation}
and
\begin{equation}
	\label{}
	\overset{(6)}{K}_0=\frac{N^3}{16}+\frac{LN}{4}+\frac{Q}{2}-\frac{1}{2}\bigg(L_k-\overset{(3)}{K}_k\bigg)\bigg(L_k-\overset{(3)}{K}_k\bigg).
\end{equation}

\section{Gauge condition}
\label{gauge}
The gauge condition we use for the metric is the harmonic gauge
\begin{equation}
	\label{hg}
	(\sqrt{-g}g^{\mu\nu})_{,\nu}=0,
\end{equation}
which reads
\begin{eqnarray}
	\label{hgi}
	\mathcal{F}^i&\equiv&\epsilon^2\bigg(\frac{1}{2}H_{,i}-\frac{1}{2}N_{,i}-H_{ik,k}\bigg)\nonumber\\
	&&+\epsilon^4\bigg(-\frac{1}{2}NN_{,i}+\frac{1}{2}H_{ik}N_{,k}+H_{il}H_{lk,k}+H_{il,k}H_{lk}-\frac{1}{2}H_{ik}H_{,k}\nonumber\\
	&&\phantom{+\epsilon^4\bigg(}-\frac{1}{2}H_{lk}H_{lk,i}-\frac{1}{2}L_{,i}+L_{i,t}-Q_{ik,k}+\frac{1}{2}Q_{kk,i}\bigg)\nonumber\\
	& = & \mathcal{O}(\epsilon^6),
\end{eqnarray}
and
\begin{eqnarray}
	\label{hgt}
	\mathcal{F}^0&\equiv&\epsilon^3\bigg(L_{k,k}-\frac{1}{2}N_{,t}-\frac{1}{2}H_{,t}\bigg)\nonumber\\
	&&+\epsilon^5\bigg(-\frac{1}{2}NH_{,t}-NN_{,t}+\frac{1}{2}H_{lk}H_{lk,t}-H_{lk}L_{l,k}+NL_{k,k}+\frac{1}{2}L_kN_k\nonumber\\
	&&\phantom{+\epsilon^5\bigg(}+\frac{1}{2}L_kH_{,k}-L_lH_{lk,k}-\frac{1}{2}L_{,t}-\frac{1}{2}Q_{kk,t}+Q_{k,k}\bigg)\nonumber\\
	& = & \mathcal{O}(\epsilon^7).
\end{eqnarray}
In addition, the divergence of the vector field equation (\ref{aevfe}), which is
\begin{equation}
	\label{csvf}
	\sum_{i=1}^{5}\Xi^{\alpha}_{(i);\alpha}=0,
\end{equation}
can be applied  to simplify our mathematical deduction \cite{TEGP}, whose expression will be given in Sec. \ref{LiK3i} in detail. 

\section{Second order post-Newtonian Approximation}

\subsection{Newtonian limit}
\label{newtonian}
The leading terms of $g_{00}$ and $K_0$ show the Newtonian limit of ae-theory. With Eqs. (\ref{lambdaexp}) and (\ref{Kt2}), $R_{00}$ to the order $\mathcal{O}(\epsilon^2)$ yields
\begin{equation}
	\label{R00e2}
	\Delta N=-8\pi \mathcal{G} \sigma,
\end{equation}
where $\Delta\equiv \nabla^2$ is the Laplace operator for the space coordinates $x^i$ and Newton's constant $\mathcal{G}$ is related to the constant $G$ by
\begin{equation}
	\label{}
	\mathcal{G}=\frac{G}{1-\frac{1}{2}c_{14}}.
\end{equation}
Here, we use a notation like $c_{ijk}$ for $c_i+c_j+c_k$, for example $c_{123}\equiv c_1+c_2+c_3$. Obviously, $N$ is twice of usual Newtonian potential $U$, which is given by the standard Poisson integral
\begin{equation}
	\label{}
	U=\Delta^{-1}\{-4\pi \mathcal{G} \sigma\}\equiv \mathcal{G}\int\frac{\sigma(\bm{x}',t)}{|\bm{x}-\bm{x}'|}\mathrm{d}^3x'.
\end{equation}

\subsection{First order post-Newtonian approximation}
\label{pna}
Following Chandrasekhar's approach \cite{Chandrasekhar1965, Chandrasekhar1969}, we look for the solution of the field equations in the form of Taylor expansion with respect to the parameter $\epsilon$. The solutions of the metric $g_{\mu \nu}$ and the vector filed $K_{\mu}$ are as follows.

\subsubsection{$H_{ij}$}
From the field equations of $R_{ij}$ with harmonic gauge, we can easily have
\begin{equation}
	\label{LapHij}
	\Delta H_{ij}=-8\pi \mathcal{G} \sigma \delta_{ij}.
\end{equation}
As $H_{ij}$ is solved in an isotropic form, it brings a lot of convenience into subsequent works. The harmonic gauge and the covariant divergence of the field equation for the vector field, Eq. (\ref{csvf}), become quite simple in 1PN approximation.

\subsubsection{$L_i$ and $\overset{(3)}{K}_i$}
\label{LiK3i}
Expanding the field equation of $R_{0i}$ and $K_i$ to $\mathcal{O}(\epsilon^3)$ with gauge (\ref{hgt}), we obtain
\begin{equation}
	\label{R0ie3}
	(1-c_{13})\Delta L_i+c_{13}\Delta\overset{(3)}{K}_i-c_{123}N_{,it}+(c_{2}+c_{123})\overset{(3)}{K}_{k,ki}=16\pi G \sigma_i,
\end{equation}
and 
\begin{equation}
	\label{Kie3}
	-\frac{1}{2}c_{13}\Delta L_i+c_1\Delta\overset{(3)}{K}_i+c_{23}\overset{(3)}{K}_{k,ki}-\frac{1}{2}(c_{23}-c_4)N_{,it}=0.
\end{equation}
To solve $\Delta L_i$ and $\Delta\overset{(3)}{K}_i$ in above equations, we use Eq. (\ref{csvf}) to eliminate the terms of $\overset{(3)}{K}_{k,ki}$. With the help of harmonic gauge and the results obtained previously, Eq. (\ref{csvf}) can be written down as
\begin{equation}
	\label{}
	\epsilon^3\Delta\bigg[\overset{(3)}{K}_{k,k}-\bigg(\frac{3}{2}-\frac{c_{12}+c_{24}}{2c_{123}}\bigg)N_{,t}\bigg]=\mathcal{O}(\epsilon^5).
\end{equation}
Therefore, we can solve $L_i$ and $\overset{(3)}{K}_i$ as
\begin{equation}
	\label{LapLi}
	\Delta L_i  = \frac{16c_1(2-c_{14})}{c_3^2+2c_1-c_1^2} \pi \mathcal{G} \sigma_i+\frac{2(c_3^2+c_1c_4)}{c_3^2+2c_1-c_1^2}N_{,it},
\end{equation}
and
\begin{equation}
	\label{LapKi3}
	\Delta\overset{(3)}{K}_i  = \frac{8c_{13}(2-c_{14})}{c_3^2+2c_1-c_1^2} \pi \mathcal{G} \sigma_i+C^{\overset{(3)}{K}_i}_{N_{,it}}N_{,it},
\end{equation}
where $C^{\overset{(3)}{K}_i}_{N_{,it}}$ is a constant. All constants with the form of $C^X_{Y_{,\mu\nu}}$ are given in Appendix \ref{cxy}.

\subsubsection{$L$}
As above, we expand the field equation of $R_{00}$ to $\mathcal{O}(\epsilon^4)$ and solve $L$ as
\begin{equation}
	\label{}
	\Delta L  =  -\frac{1}{2}\Delta N^2+C^L_{N_{,tt}}N_{,tt}.
\end{equation}

\subsubsection{Summary of PPN parameters}
\label{sumppn}
In 1PN approximation, the metric (\ref{PNmetric1})-(\ref{PNmetric3}) is equivalent to the PPN metric \cite{Will1972,TEGP} under a trivial gauge transformation. The transformation between our reference system $(t,x^i)$ and the PPN reference system in the standard PN gauge $(t_{\mathrm{PN}},x^i_{\mathrm{PN}})$ reads
\begin{eqnarray}
	\label{}
	t_{\mathrm{PN}} & = & t+\epsilon^4\lambda_1\chi_{,t}+\mathcal{O}(\epsilon^6),\\
	x^i_{\mathrm{PN}} & = & x^i,
\end{eqnarray}
where
\begin{equation}
	\label{}
	\lambda_1=-\frac{1}{1-\frac{1}{2}c_{14}}\bigg[1-\frac{7}{2}c_1-\frac{9}{2}c_2-\frac{3}{2}c_3-2c_4+\frac{1}{2c_{123}}(c_2+c_{12}+c_{1234})^2\bigg],
\end{equation}
and $\chi$ is the superpotential defined by
\begin{equation}
	\label{}
	\chi=\frac{1}{2}\mathcal{G}\int\sigma(\bm{x}',t)|\bm{x}-\bm{x}'|\mathrm{d}^3x'+\mathcal{O}(\epsilon^2),
\end{equation}
so that
\begin{equation}
	\label{}
	\Delta\chi=\frac{1}{2}N.
\end{equation}
After transformation, 5 PPN parameters are
\begin{eqnarray}
	\label{}
	\gamma & = & 1,\\
	\beta & = & 1,\\
	\xi & = & 0,\\
	\alpha_1 & = & -\frac{8(c_1c_4+c_3^2)}{c_3^2+2c_1-c_1^2},
\end{eqnarray}
and
\begin{equation}
	\label{}
	\alpha_2  =  \frac{(2c_{13}-c_{14})^2}{c_{123}(2-c_{14})}-\frac{12c_3c_{13}+2c_1c_{14}(1-2c_{14})+(c_1^2-c_3^2)(4-6c_{13}+7c_{14})}{(2-c_{14})(c_3^2+2c_1-c_1^2)}.
\end{equation}
Other 5 conservation law parameters, $\alpha_3$, $\zeta_1$, $\zeta_2$, $\zeta_3$ and $\zeta_4$, are all zero due to the theorem of Lee, et al. \cite{Lee1974}.  These results perfectly match previous works \cite{Eling2004,Graesser2005,Foster2006a}.

\subsection{Second order post-Newtonian approximation}

Following above procedures, we can obtain 2PN approximation of ae-theory. The $\mathcal{O}(\epsilon^4)$ term in $g_{ij}$ is solved as
\begin{eqnarray}
	\label{}
	\Delta Q_{ij} & = & -\bigg(1-\frac{1}{2}c_{14}\bigg)N_{,i}N_{,j}+c_{13}\bigg(L_{i,jt}+L_{j,it}-\overset{(3)}{K}_{i,jt}-\overset{(3)}{K}_{j,it}\bigg)\nonumber\\
	&&-8(2-c_{14})\pi \mathcal{G} \sigma_{ij}+\delta_{ij}\bigg[+\frac{1}{2}\Delta N^2+8(2-c_{14})\pi \mathcal{G} \sigma_{kk}+C^{Q_{ij}}_{N_{,tt}}N_{,tt}\bigg].
\end{eqnarray}

As we can see, $Q_{ij}$ no longer keeps isotropic as $H_{ij}$. To succeed in the convention and convenience in 1PN, we will try to transfer $Q_{ij}$ into an isotropic form as possible as we can in next section for the application in the light propagation model.

Before solving $Q_i$ and $\overset{(5)}{K}_i$, we need Eq. (\ref{csvf}) to attain to $\mathcal{O}(\epsilon^5)$. It gives
\begin{eqnarray}
	\label{withA}
	&&\epsilon^3\Delta\bigg[\overset{(3)}{K}_{k,k}-\bigg(\frac{3}{2}-\frac{c_{12}+c_{24}}{2c_{123}}\bigg)N_{,t}\bigg]\nonumber\\
	&&+\epsilon^5\Delta\bigg\{\bigg[\frac{1}{c_{123}}\bigg(\frac{1}{2}c_{1}-\frac{3}{4}c_2-\frac{1}{4}c_3+\frac{3}{4}c_4\bigg)-1\bigg]NN_{,t}+\frac{1}{c_{123}}\bigg(\frac{5}{8}c_{13}+\frac{1}{2}c_2\bigg)NL_{k,k}\nonumber\\
	&&\phantom{+\epsilon^5\Delta\bigg\{}+\frac{1}{2}L_kN_{,k}-\bigg(1-\frac{c_4}{4c_{123}}\bigg)N\overset{(3)}{K}_{k,k}-\frac{c_{13}}{2c_{123}}Q_{kk,t}+\frac{1}{2}\bigg(\frac{c_{14}}{c_{123}}-1\bigg)L_{,t}\nonumber\\
	&&\phantom{+\epsilon^5\Delta\bigg\{}+\overset{(5)}{K}_{k,k}+\mathcal{A}\bigg\}=\mathcal{O}(\epsilon^7),
\end{eqnarray}
where $\Delta \mathcal{A}$ satisfies a Poisson's equation given in Appendix \ref{cxy}.
After eliminating the terms related to $\overset{(5)}{K}_{k,k}$ in the field equations, we can solve $Q_i$ and $\overset{(5)}{K}_i$ as
\begin{eqnarray}
	\label{}
	\Delta Q_i & = & \frac{16c_1(2-c_{14})}{c_3^2-c_1^2+2c_1}\pi \mathcal{G} N \sigma_i-\frac{8(c_1^2+c_3^2+2c_1c_4-2c_1)}{c_3^2-c_1^2+2c_1}\pi \mathcal{G} \sigma L_i\nonumber\\
	&&+\frac{16(c_1c_4+c_3^2)}{c_3^2-c_1^2+2c_1}\pi \mathcal{G} \sigma \overset{(3)}{K}_i+C^{Q_i}_{N_{,i}N_{,t}}N_{,i}N_{,t}+C^{Q_i}_{NN_{,it}}NN_{,it}\nonumber\\
	&&+\frac{2c_1}{c_3^2-c_1^2+2c_1}N_{,ik}L_k-2N_{,k}L_{k,i}+\frac{c_{13}c_{14}}{c_3^2-c_1^2+2c_1}N_{,k}L_{i,k}\nonumber\\
	&&-\frac{2c_1+c_{13}c_{14}}{c_3^2-c_1^2+2c_1}L_kN_{,ki}+L_{i,tt}+\frac{2c_1c_{14}}{c_3^2-c_1^2+2c_1}L_{,it}\nonumber\\
	&&+\frac{c_{13}c_{14}}{c_3^2-c_1^2+2c_1}\overset{(3)}{K}_kN_{,ki}+\frac{c_{14}(c_1-c_3)}{c_3^2-c_1^2+2c_1}\overset{(3)}{K}_{k,i}N_{,k}-\frac{2c_1c_{14}}{c_3^2-c_1^2+2c_1}\overset{(3)}{K}_{i,k}N_{,k}\nonumber\\
	&&-\frac{2c_{13}c_{14}}{c_3^2-c_1^2+2c_1}\overset{(3)}{K}_{i,tt}+\frac{c_3^2-c_1^2}{c_3^2-c_1^2+2c_1}Q_{kk,it}-\frac{2(c_3c_{23}-c_1c_{12})}{c_3^2-c_1^2+2c_1}\mathcal{A}_{,i},
\end{eqnarray}
and
\begin{eqnarray}
	\label{}
	\Delta\overset{(5)}{K}_i & = & \frac{12c_{13}(2-c_{14})}{c_3^2-c_1^2+2c_1}\pi \mathcal{G} N\sigma_i+\frac{8c_{13}(2-c_{14})}{c_3^2-c_1^2+2c_1}\pi \mathcal{G} \sigma L_i\nonumber\\
	&&+\frac{4(c_1^2+2c_1c_{34}+c_3^2+2c_3c_4-4c_3-2c_1)}{c_3^2-c_1^2+2c_1}\pi \mathcal{G} \sigma \overset{(3)}{K}_i\nonumber\\
	&&+C^{\overset{(5)}{K}_i}_{N_{,i}N_{,t}}N_{,i}N_{,t}+C^{\overset{(5)}{K}_i}_{NN_{,it}}NN_{i,t}+\frac{c_{13}}{c_3^2-c_1^2+2c_1}N_{,ik}L_k\nonumber\\
	&&-\frac{1}{2}N_{,k}L_{k,i}+\frac{c_{14}(c_{13}-1)}{c_3^2-c_1^2+2c_1}N_{,k}L_{i,k}-\frac{c_{13}^2+2c_{13}(c_4+1)-2c_4}{2(c_3^2-c_1^2+2c_1)}L_kN_{,ki}\nonumber\\
	&&+C^{\overset{(5)}{K}_i}_{L_{,it}}L_{,it}+\frac{c_{13}(c_4+c_{134})-2c_4}{2(c_3^2-c_1^2+2c_1)}\overset{(3)}{K}_kN_{,ki}+\frac{3c_1^2-3c_3^2-4c_1+2c_4}{2(c_3^2-c_1^2+2c_1)}\overset{(3)}{K}_{k,i}N_{,k}\nonumber\\
	&&-\frac{c_{13}(2c_1+c_4-c_3)-2c_1}{c_3^2-c_1^2+2c_1}\overset{(3)}{K}_{i,k}N_{,k}-\frac{2c_{14}(c_{13}-1)}{c_3^2-c_1^2+2c_1}\overset{(3)}{K}_{i,tt}\nonumber\\
	&&+\frac{c_{13}(c_3^2-c_1^2-2c_{23})}{2c_{123}(c_3^2-c_1^2+2c_1)}Q_{kk,it}+\frac{c_1^2-c_3^2+2c_{23}}{c_3^2-c_1^2+2c_1}\mathcal{A}_{,i}.
\end{eqnarray}
Similarly, $Q$ can be solved as
\begin{eqnarray}
	\label{}
	\Delta Q & = & 8\pi \mathcal{G} N^2 \sigma+16\pi \mathcal{G} L \sigma -\frac{16(c_3^2+c_1^2+2c_1c_4-2c_1)}{c_3^2-c_1^2+2c_1}\pi \mathcal{G} L_k\sigma_k\nonumber\\
	&&+\frac{32(c_1c_4+c_3^2)}{c_3^2-c_1^2+2c_1}\pi \mathcal{G} \overset{(3)}{K}_k\sigma_k-16\pi \mathcal{G} N\sigma_{kk}+C^{Q}_{NM_{,tt}}NN_{,tt}\nonumber\\
	&&+C^{Q}_{N_{,t}N_{,t}}N_{,t}N_{,t}-N_{,k}L_{k,t}+C^Q_{N_{,kt}L_k}N_{,kt}L_k-\frac{3}{2}N_{,k}L_{,k}\nonumber\\
	&&+N_{,k}Q_{kl,l}+N_{,kl}Q_{kl}-\frac{1}{2}N_{,l}Q_{kk,l}-\frac{2(1-c_{13})}{2-c_{14}}L_{k,l}L_{l,k}\nonumber\\
	&&+\frac{2(1+c_3-c_4)}{2-c_{14}}L_{l,k}L_{l,k}+\frac{1}{2-c_{14}}\bigg[2+\frac{c_{14}(c_{12}+c_{24})}{c_{123}}\bigg]L_{,tt}\nonumber\\
	&&+C^Q_{N_{,kt}\overset{(3)}{K}_k}N_{,kt}\overset{(3)}{K}_k-\frac{2(c_1+2c_3-c_4)}{2-c_{14}}\bigg(L_{k,l}\overset{(3)}K_{l,k}+L_{l,k}\overset{(3)}{K}_{l,k}\bigg)\nonumber\\
	&&-\frac{2(c_{23}-c_4)}{2-c_{14}}L_k\overset{(3)}{K}_{l,lk}+\frac{2(c_1-c_4)}{2-c_{14}}\overset{(3)}{K}_{l,k}\overset{(3)}{K}_{k,l}+\frac{4c_3}{2-c_{14}}\overset{(3)}{K}_{l,k}\overset{(3)}{K}_{l,k}\nonumber\\
	&&-\frac{c_{13}}{2-c_{14}}\bigg(1+\frac{c_{12}+c_{24}}{c_{123}}\bigg)Q_{kk,tt}+\frac{2(c_{12}+c_{24}+c_{123})}{2-c_{14}}\mathcal{A}_{,t}.
\end{eqnarray}

\subsection{Verification of the solutions of metric and vector field}
One way to verify the solutions obtained above is to check the gauge condition. Inserting the metric coefficients into the harmonic gauge, we have
\begin{equation}
	\label{}
	\Delta \mathcal{F}^0  = \epsilon^3\frac{32c_1\pi G(1+\epsilon^2N)}{c_3^2+2c_1-c_1^2}\bigg[\sigma_{,t}+\sigma_{k,k}+\epsilon^2\bigg(\frac{1}{2}\sigma N_{,t}-\sigma_{kk,t}\bigg)\bigg]+\mathcal{O}(\epsilon^7),
\end{equation}
and
\begin{equation}
	\label{}
	\Delta \mathcal{F}^i=16\epsilon^4\pi G  \bigg(\sigma_{i,t}+\sigma_{ik,k}-\frac{1}{2}\sigma N_{,i}\bigg)+\mathcal{O}(\epsilon^6),
\end{equation}
which are equivalent to the equation of motion $T^{\mu\nu}_{;\nu}=0$.

Another approach as the most reliable way to verify the results is to substitute the metric and vector 2PN expansion coefficients into the field equations and check. Our results pass this examination.

\subsection{The 2PN parameters $\iota$}
The 2PN metric of ae-theory shows a picture of second order PPN (2PPN) formalism. Previous works of 2PPN formalism focus on a many-body Lagrangian \cite{Benacquista1988,Benacquista1992,Nordtvedt1993}. Whereas we have not restricted the matter in our model, we are not going to compare our results with theirs in this paper. Despite of applying a ``theory-dependent'' approach, several works also obtain some 2PN parameters. Although there are many parameters regarding the effects of preferred frame, here we concentrate only on the parameter that represents the third order nonlinearity of Newtonian potential in $g_{00}$ only here. Damour and Esposito-Far\`{e}se calculate two 2PN parameters $\varepsilon$ and $\zeta$ in a multiscalar-tensor theory, and they find that the possible 2PN \emph{deviations} from GR, $\delta g^{\mathrm{DE}}_{00}\equiv g^{\mathrm{DE}}_{00}-g^{\mathrm{GR}}_{00}$,  are given by \cite{Damour1992, Damour1996}
\begin{equation}
	\label{}
	\Delta \delta g^{\mathrm{DE}}_{00}=\frac{\varepsilon}{3c^6}\Delta U^3-\frac{\varepsilon}{c^6}4\pi G\sigma U^2+\mathcal{O}\bigg(\frac{\gamma}{c^6},\frac{\beta}{c^6}\bigg)+\mathcal{O}\bigg(\frac{1}{c^8}\bigg),
\end{equation}
where only one scalar field is involved, $U$ is the Newtonian potential, $\varepsilon$ measures how much the third order nonlinearity there is in the superposition law for gravity and $\zeta$ is no longer an independent 2PN parameter with $\zeta=\zeta(\gamma,\beta)$. Recently, Xie \textit{et al.}\cite{Xie2007} find a 2PN parameter $\iota$ in a scalar-tensor theory (STT) with a intermediate range force, in which $\iota$ in $\delta g^{\mathrm{STT}}_{00} \equiv g^{\mathrm{STT}}_{00}-g^{\mathrm{GR}}_{00}$ has the form as
\begin{equation}
	\label{}
	\Delta \delta g^{\mathrm{STT}}_{00}=-\frac{\iota}{c^6}U\Delta U^2+\mathcal{O}\bigg(\frac{\gamma}{c^6},\frac{\beta}{c^6}\bigg)+\mathcal{O}\bigg(\frac{1}{c^8}\bigg).
\end{equation}
With the help of $\Delta U=-4\pi G \sigma$ and $\Delta U^3=3U\Delta U^2-3U^2\Delta U$, $\iota$ and $\varepsilon$ are equivalent and they represent the third order nonlinearity in 2PN $g_{00}$, which is totally different from the terms due to the combinations of 1PN terms, $\mathcal{O}(\gamma/c^6,\beta/c^6)$. In the case of ae-theory, $\iota=0$, which makes ae-theory no differences with GR in the parameters of $\gamma$, $\beta$ and $\iota$ (see Tab. \ref{tab2}). In this table, $\omega_0$, $\omega_1$ and $\omega_2$ are constants, coming from the expansion of the coupling function in the STT with an intermediate-range force (see \cite{Xie2007} for details). But even so, experiments, especially the deep space laser ranging missions, still can test ae-theory by its unique effects that deviate from GR, such as the 2PN light deflection angle.
\begin{table}
	\centering
	\caption{\label{tab2}Summary of the parameters.}
	\begin{tabular}{c|l|c|c|c}
		\hline
		Parameter & What it measures, relative to GR & Value in GR & Value in STT \cite{Xie2007} & Value in ae-theory\\
		\hline
		$\gamma$ & How much space curvature $(g_{ij})$ & $1$ & $\frac{\omega_0+1}{\omega_0+2}$ & $1$\\
		&  is produced by unit rest mass?\cite{mtw}& & &\\
		$\beta$ & How much the second order nonlinearity   & $1$ & $1+\frac{\omega_1}{(2\omega_0+3)(2\omega_0+4)^2}$ & $1$\\
		&is there in the superposition law& & &\\
		&for gravity $(g_{00})$?\cite{mtw} & & &\\
		\hline
		$\iota$ & How much the third order nonlinearity& 0 & $\frac{\omega_2}{2(3+2\omega_0)(\omega_0+2)^3}$ & $0$\\
		&is there in the superposition law&  & &\\
		&  for gravity $(g_{00})$? & & &\\
		\hline
	 \end{tabular}
\end{table}

\section{A 2PN metric for light propagation in the solar system}
Future deep space laser ranging missions such as Laser Astrometric Test of Relativity (LATOR) \cite{LATOR}, and Astrodynamics Space Test of Relativity (ASTROD) \cite{ASTROD1}, together with astrometry missions such as Global Astrometric Interferometer for Astrophysics (GAIA) \cite{GAIA} and Space Interferometry Mission (SIM) \cite{SIM} will be able to test relativistic gravity to an unprecedented level of accuracy in the solar system. Those missions will enable us to test relativistic gravity to $10^{-6}-10^{-9}$, and will require 2PN approximation of relevant theories of gravity, including metric coefficients, equations of motion and equations of light ray.

Hence, in this section, we discuss a 2PN metric for light propagation in the solar system. Considering a practical model, we just study a situation of $N$ point masses in a global frame as the first step. And we impose a constraint on the metric, that is, after ignoring all the planets in the solar system, the spatial part of the metric, $g_{ij}$, should be isotropic after a coordinate transformation.

\subsection{A coordinate transformation}
A coordinate transformation
\begin{equation}
	\label{}
	\bar{x}^{\mu}  = x^{\mu}+\epsilon^4\xi^{\mu}(x^{\alpha}),
\end{equation}
and
\begin{equation}
	\label{}
	{x}^{\mu}  = \bar{x}^{\mu}-\epsilon^4\xi^{\mu}(\bar{x}^{\alpha}),
\end{equation}
where $\xi^{\mu}\sim \mathcal{O}(1)$, changes the metric to
\begin{eqnarray}
	\label{gbar}
	\bar{g}_{\mu\nu} & = & \frac{\partial x^{\rho}}{\partial \bar{x}^{\mu}}\frac{\partial x^{\lambda}}{\partial \bar{x}^{\nu}}g_{\rho \lambda}(x^\alpha)\nonumber\\
	& = & g_{\mu\nu}(x^{\alpha})-\epsilon^4g_{\mu\rho}\xi^{\rho}_{,\nu}(x^{\alpha})-\epsilon^4g_{\nu\rho}\xi^{\rho}_{,\mu}(x^{\alpha})+\mathcal{O}(\epsilon^8).
\end{eqnarray}
When the time component is chosen to be fixed ($\bar{t}=t$), we obtain the metric to 2PN order
\begin{eqnarray}
	\label{bgij}
	\bar{g}_{ij} & = & g_{ij}(x^{\alpha})-\epsilon^4\xi^i_{,j}(x^{\alpha})-\epsilon^4\xi^{j}_{,i}(x^{\alpha})+\mathcal{O}(\epsilon^5),\\
	\label{bg0i}
	\bar{g}_{0i} & = & g_{0i}(x^{\alpha})-\epsilon^5 \xi^{i}_{,t}(x^{\alpha})+\mathcal{O}(\epsilon^6),\\
	\label{bg00}
	\bar{g}_{00} & = & g_{00}(x^{\alpha})+\mathcal{O}(\epsilon^7).
\end{eqnarray}
Then, $x^{\alpha}$ in the right hand side of above three equations need to be replaced with $\bar{x}^{\alpha}$. We must also transform the functional integrals over $x^{k'}$ that appear in $g_{\mu\nu}$ into integrals over $\bar{x}^{k'}$. The only place where this changes anything is in $g_{00}=-1+\epsilon^2N(\bm{x},t)+\mathcal{O}(\epsilon^4)$, where
\begin{equation}
	\label{}
	N(\bm{x},t) =2\mathcal{G}\int \frac{\sigma(\bm{x}',t)}{|\bm{x}-\bm{x}'|}\mathrm{d}^3x'.
\end{equation}
Like Eq. (\ref{gbar}), the transferred energy-momentum tensor $\bar{T}^{\mu\nu}$ is
\begin{eqnarray}
	\label{}
	\bar{T}^{\mu\nu} & = & \frac{\partial \bar{x}^{\mu}}{\partial x^{\rho}}\frac{\partial \bar{x}^{\nu}}{\partial x^{\lambda}}T^{\rho \lambda}\nonumber\\
	& = & T^{\mu\nu}+\epsilon^4T^{\mu\rho}\xi^{\nu}_{,\rho}+\epsilon^4T^{\nu\rho}\xi^{\mu}_{,\rho}+\mathcal{O}(\epsilon^8).
\end{eqnarray}
According to the definition of $\sigma$ (\ref{defsigma}), it can be obtained that 
\begin{eqnarray}
	\label{bsigma}
	\bar{\sigma} & = & \bar{T}^{00}+\bar{T}^{ss}\nonumber\\
	& = & T^{00}+T^{ss}+2\epsilon^4T^{\rho s}\xi^{s}_{,\rho}\nonumber\\
	& = & \sigma +\mathcal{O}(\epsilon^6).
\end{eqnarray}
Furthermore, the difference between the volume element $\mathrm{d}^3x'$ and $\mathrm{d}^3\bar{x}'$ is a Jacobian determinant that
\begin{eqnarray}
	\label{}
	\mathrm{d}^3{x}' & = & \bigg|\frac{\partial {\bm{x}}'}{\partial \bar{\bm{x}}'}\bigg|\mathrm{d}^3\bar{x}'\nonumber\\
	& = & \mathrm{d}^3\bar{x}'[1-\epsilon^4\bar{\nabla}' \cdot \bar{\bm{\xi}}'+\mathcal{O}(\epsilon^8)].
\end{eqnarray}
where $\bar{\bm{\xi}}=\bm{\xi}(\bar{\bm{x}})$. We also have
\begin{eqnarray}
	\label{}
	\frac{1}{|\bm{x}-\bm{x}'|} & = & \frac{1}{|\bar{\bm{x}}-\bar{\bm{x}}'-\epsilon^4(\bar{\bm{\xi}}-\bar{\bm{\xi}}')|} \nonumber\\
	& = & \frac{1}{|\bar{\bm{x}}-\bar{\bm{x}}'|}+\epsilon^4\frac{(\bar{\bm{x}}-\bar{\bm{x}}')\cdot(\bar{\bm{\xi}}-\bar{\bm{\xi}}')}{|\bar{\bm{x}}-\bar{\bm{x}}'|^3}+\mathcal{O}(\epsilon^8).
\end{eqnarray}
Thus, we put these relations together and have
\begin{eqnarray}
	\label{}
	N(\bm{x},t) & = & \bar{N}(\bar{\bm{x}},\bar{t})-2\epsilon^4\mathcal{G}\int\frac{\bar{\sigma}'\bar{\nabla}'\cdot\bar{\bm{\xi}}'}{|\bar{\bm{x}}-\bar{\bm{x}}'|}\mathrm{d}^3\bar{x}'\nonumber\\
	&&+2\epsilon^4\mathcal{G}\int\frac{\bar{\sigma}'(\bar{\bm{x}}-\bar{\bm{x}}')\cdot(\bar{\bm{\xi}}-\bar{\bm{\xi}}')}{|\bar{\bm{x}}-\bar{\bm{x}}'|^3}\mathrm{d}^3\bar{x}'+\mathcal{O}(\epsilon^6).
\end{eqnarray}
Finally, we obtain the metric after transformation
\begin{eqnarray}
	\label{bgij2}
	\bar{g}_{ij} & = & g_{ij}(\bar{x}^{\alpha})-\epsilon^4\xi^i_{,j}(\bar{x}^{\alpha})-\epsilon^4\xi^{j}_{,i}(\bar{x}^{\alpha})+\mathcal{O}(\epsilon^5),\\
	\label{bg0i2}
	\bar{g}_{0i} & = & g_{0i}(\bar{x}^{\alpha})-\epsilon^5 \xi^{i}_{,t}(\bar{x}^{\alpha})+\mathcal{O}(\epsilon^6),\\
	\label{bg002}
	\bar{g}_{00} & = & g_{00}(\bar{x}^{\alpha})-2\epsilon^6\mathcal{G}\bar{\Delta}^{-1}\{-4\pi \bar{\sigma} \bar{\nabla} \cdot \bar{\bm{\xi}}\}\nonumber\\
	&&+2\epsilon^6\mathcal{G}\bar{\nabla} \cdot \bar{\Delta}^{-1}\{-4\pi \bar{\sigma} \bar{\bm{\xi}}\}\nonumber\\
	&&+2\epsilon^6\mathcal{G}\int\frac{\bar{\sigma}'(\bar{\bm{x}}-\bar{\bm{x}}')\cdot\bar{\bm{\xi}}}{|\bar{\bm{x}}-\bar{\bm{x}}'|^3}\mathrm{d}^3\bar{x}'+\mathcal{O}(\epsilon^7).
\end{eqnarray}
So if $\xi^i=c_{13}\Delta^{-1}(L_{i,t}-\overset{(3)}{K}_{i,t})$, the term of $c_{13}\Delta^{-1}(L_{i,jt}+L_{j,it}-\overset{(3)}{K}_{i,jt}-\overset{(3)}{K}_{j,it})$ can be eliminated and the 1PN parts of metric keep unchanged. With the help of this transformation, the metric for light can be changed into a simpler form.

\subsection{A 2PN light propagation metric of $N$ point masses}
In light propagation, we can cut off the full 2PN metric to
\begin{eqnarray}
	\label{}
	g_{00} & = & -1+\epsilon^2N+\epsilon^4L+\mathcal{O}(\epsilon^5)\\
	g_{0i} & = & \epsilon^3L_i+\mathcal{O}(\epsilon^5),\\
	g_{ij} & = & \delta_{ij} +\epsilon^2H_{ij}+\epsilon^4Q_{ij}+\mathcal{O}(\epsilon^5).
\end{eqnarray}
If we consider the solar system as a $N$-body problem of non-spinning point masses for simplicity, we follow the notation adopted by \cite{Blanchet1998,Faye2006} and use the matter stress-energy tensor
\begin{equation}
	\label{}
	c^2T^{\mu\nu}(\bm{x},t)=\sum_a \mu_a(t)v^{\mu}_a(t)v^{\nu}_a(t)\delta(\bm{x}-\bm{y}_a(t)),
\end{equation}
where $\delta$ denotes the three-dimensional Dirac distribution, the trajectory of the $a$th mass is represented by $\bm{y}_a(t)$, the coordinate velocity of the $a$th body are $\bm{v}_a=\mathrm{d}\bm{y}_a(t)/\mathrm{d}t$ and $v_a^{\mu}\equiv(c,\bm{v}_a)$ and $\mu_a$ denotes an effective time-dependent mass of the $a$th body defined by
\begin{equation}
	\label{}
	\mu_a=\bigg(\frac{m_a}{\epsilon\sqrt{gg_{\rho\lambda}v_a^{\rho}v_a^{\lambda}}}\bigg)_a,
\end{equation}
where $(\cdot)_a$ means evaluation at the $a$th body and $m_a$ being the constant Schwarzschild mass. Another useful notation is
\begin{equation}
	\label{}
	\tilde{\mu}_a(t)=\mu_a(t)(1+\epsilon^2v_a^2),
\end{equation}
where $v_a^2=\bm{v}_a^2$. Both $\mu_a$ and $\tilde{\mu}_a$ reduce to the Schwarzschild mass at Newtonian order: $\mu_a=m_a+\mathcal{O}(\epsilon^2)$ and $\tilde{\mu}_a=m_a+\mathcal{O}(\epsilon^2)$. Then the mass, current and stress densities (\ref{defsigma}-\ref{defsigmaij}) for the $N$ point masses read
\begin{eqnarray}
	\label{}
	\sigma & = & \sum_a\tilde{\mu}_a\delta(\bm{x}-\bm{y}_a),\\
	\sigma_i & = & \sum_a \mu_a v_a^i\delta(\bm{x}-\bm{y}_a),\\
	\sigma_{ij} & = & \sum_a \mu_a v_a^iv_a^j\delta(\bm{x}-\bm{y}_a).
\end{eqnarray}
Therefore, we can work out $N$ and $H_{ij}$ quickly,
\begin{eqnarray}
	\label{}
	N  & = & 2\Delta^{-1}\{-4\pi \mathcal{G} \sigma\}\nonumber\\
	& = & 2\sum_a\frac{\mathcal{G}m_a}{r_a}\bigg\{1+\epsilon^2\bigg[+\frac{3}{2}v_a^2-\sum_{b\neq a} \frac{\mathcal{G}m_b}{r_{ab}}\bigg]+\mathcal{O}(\epsilon^4)\bigg\},
\end{eqnarray}
\begin{equation}
	\label{}
	H_{ij}=2\delta_{ij}\sum_a\frac{\mathcal{G}m_a}{r_a}\bigg\{1+\epsilon^2\bigg[+\frac{3}{2}v_a^2-\sum_{b\neq a} \frac{\mathcal{G}m_b}{r_{ab}}\bigg]+\mathcal{O}(\epsilon^4)\bigg\},
\end{equation}
by the relation of $\tilde{\mu}_a$ that
\begin{eqnarray}
	\label{}
	\tilde{\mu}_a & = &  m_a\bigg\{1+\epsilon^2\bigg[\bigg(N-\frac{1}{2}H\bigg)_a+\frac{3}{2}v_a^2\bigg]+\mathcal{O}(\epsilon^4)\bigg\}\nonumber\\
	& = & m_a\bigg\{1+\epsilon^2\bigg[+\frac{3}{2}v_a^2-\sum_{b\neq a} \frac{\mathcal{G}m_b}{r_{ab}}\bigg]+\mathcal{O}(\epsilon^4)\bigg\}.
\end{eqnarray}
where $r_a=|\bm{x}-\bm{y}_a|$ and $r_{ab}=|\bm{y}_a-\bm{y}_b|$.

In the solution of $L_i$, due to
\begin{equation}
	\label{}
	\Delta^{-1}\{-4\pi \mathcal{G} \sigma_i\}=\sum_a \int \frac{\mathcal{G}m_av_a^i\delta(\bm{z}-\bm{y}_a)}{|\bm{x}-\bm{z}|}\mathrm{d}^3z=\sum_a \frac{\mathcal{G}m_av_a^i}{r_a}+\mathcal{O}(\epsilon^2),
\end{equation}
and
\begin{eqnarray}
	\label{}
	N_{,it} & = & 2\Delta \chi_{,it}\nonumber\\
	& = & \Delta\bigg\{\mathcal{G}\sum_a \int m_a\delta(\bm{z}-\bm{y}_a)|\bm{x}-\bm{z}|_{,i}\mathrm{d}^3z+\mathcal{O}(\epsilon^2)  \bigg\}_{,t}\nonumber\\
	& = & \Delta\bigg\{\sum_a \mathcal{G}m_an_a^i+\mathcal{O}(\epsilon^2)\bigg\}_{,t}\nonumber\\
	& = & \Delta \bigg\{\sum_a \frac{\mathcal{G}m_a}{r_a}\bigg[(\bm{n}_a\cdot \bm{v}_a )n_a^i-v_a^i \bigg]+\mathcal{O}(\epsilon^2)\bigg\},
\end{eqnarray}
where
\begin{equation}
	\label{}
	\chi=\frac{1}{2}\mathcal{G}\int\sigma(\bm{x}',t)|\bm{x}-\bm{x}'|\mathrm{d}^3x'+\mathcal{O}(\epsilon^2),
\end{equation}
it can be solved as
\begin{equation}
	\label{}
	L_i=-\frac{2c_{13}^2-2c_1c_{14}+8c_1}{c_3^2-c_1^2+2c_1}\sum_a\frac{\mathcal{G}m_av_a^i}{r_a}+\frac{2(c_3^2+c_1c_4)}{c_3^2-c_1^2+2c_1}\sum_a\frac{\mathcal{G}m_a}{r_a}(\bm{n}_a\cdot \bm{v}_a )n_a^i+\mathcal{O}(\epsilon^2),
\end{equation}
where $\bm{n_a}=(\bm{x}-\bm{y}_a)/r_a$.

In $g_{00}$, with the help of
\begin{equation}
	\label{}
	|\bm{x}-\bm{y}_a|_{,t}=-\bm{n}_a\cdot\bm{v}_a,
\end{equation}
and
\begin{equation}
	\label{}
	|\bm{x}-\bm{y}_a|_{,tt}=\frac{v_a^2}{r_a}+\sum_{b\neq a}\frac{\mathcal{G}m_b}{r_{ab}^2}\bm{n}_a\cdot\bm{n}_{ab}-\frac{(n_{a}v_a)^2}{r_a},
\end{equation}
we can obtain
\begin{eqnarray}
	\label{}
	N_{,tt} & = & 2\Delta \chi_{,tt}\nonumber\\
	& = & \Delta \bigg\{ \mathcal{G}\sum_a\int m_a \delta(\bm{z}-\bm{y}_a)|\bm{x}-\bm{z}|\mathrm{d}^3\bm{z}+\mathcal{O}(\epsilon^2) \bigg\}_{,tt}\nonumber\\
	& = & \Delta \bigg\{ \mathcal{G}\sum_a m_a |\bm{x}-\bm{y}_a|_{,tt}+\mathcal{O}(\epsilon^2) \bigg\}\nonumber\\
	& = & \Delta \bigg\{ \sum_a\frac{\mathcal{G}m_a}{r_a}\bigg[ v_a^2-(n_{a}v_a)^2\bigg]+\sum_a\sum_{b\neq a}\frac{\mathcal{G}^2m_am_b}{r_{ab}^2}\bm{n}_a\cdot\bm{n}_{ab}+\mathcal{O}(\epsilon^2) \bigg\},
\end{eqnarray}
which leads to
\begin{eqnarray}
	\label{}
	L & = & -2\sum_a \frac{\mathcal{G}^2m_a^2}{r_a^2}+C^L_{N_{,tt}}\sum_a\frac{\mathcal{G}m_a}{r_a}\bigg[ v_a^2-(\bm{n}_a\cdot\bm{v}_a)^2\bigg]\nonumber\\
	&&-2\sum_a\sum_{b\neq a} \frac{\mathcal{G}^2m_am_b}{r_ar_b}+C^L_{N_{,tt}}\sum_a\sum_{b\neq a}\frac{\mathcal{G}^2m_am_b}{r_{ab}^2}\bm{n}_a\cdot\bm{n}_{ab}+\mathcal{O}(\epsilon^2),
\end{eqnarray}
where $\bm{n}_{ab}=(\bm{y}_a-\bm{y}_b)/r_{ab}$.

The quadratic part of potentials in $Q_{ij}$ can be rewritten as
\begin{eqnarray}
	\label{}
	N_{,i}N_{,j} & = & 4\sum_a \mathcal{G}^2m_a^2\bigg(\frac{1}{r_a}\bigg)_{,i}\bigg(\frac{1}{r_a}\bigg)_{,j}+4\sum_a\sum_{b\neq a} \mathcal{G}^2m_am_b\bigg(\frac{1}{r_a}\bigg)_{,i}\bigg(\frac{1}{r_b}\bigg)_{,j}\nonumber\\
	& = & \frac{1}{2}\sum_a\mathcal{G}^2m_a^2(\partial^2_{ij}+\delta_{ij}\Delta)\bigg(\frac{1}{r_a^2}\bigg)+4\sum_a\sum_{b\neq a} \mathcal{G}^2m_am_b\partial_{ai}\partial_{bj}\bigg(\frac{1}{r_ar_b}\bigg),
\end{eqnarray}
where $\partial_{ai}$ denotes the partial derivative with respect to $y_a^i$. The integral of the self-terms can be readily deduced from $\Delta (\ln r_a)=1/r_a^2$ \cite{Blanchet1998}; on the other hand, the interaction terms are obtained by
\begin{equation}
	\label{}
	\Delta \ln S_{ab}=\frac{1}{r_ar_b},
\end{equation}
where $S_{ab}\equiv r_a+r_b+r_{ab}$ \cite{Fock1959}. Consequently the first term in $\Delta Q_{ij}$ can be solved as
\begin{eqnarray}
	\label{}
	\Delta^{-1}\{N_{,i}N_{,j}\} & = & \frac{1}{2}\sum_a\mathcal{G}^2m_a^2\bigg(\partial^2_{ij}\ln r_a+\frac{\delta_{ij}}{r_a^2}\bigg)+4\sum_a\sum_{b\neq a} \mathcal{G}^2m_am_b\partial_{ai}\partial_{bj}\ln S_{ab}\nonumber\\
	& = &  \sum_a\mathcal{G}^2m_a^2\bigg(-\frac{n_a^in_a^j}{r_a^2}+\frac{\delta_{ij}}{r_a^2}\bigg)\nonumber\\
	&&+4\sum_a\sum_{b\neq a} \mathcal{G}^2m_am_b\bigg[\frac{n_{ab}^{ij}-\delta^{ij}}{r_{ab}S_{ab}}+\frac{(n_{ab}^i-n_a^i)(n_{ab}^j+n_b^j)}{S_{ab}^2}\bigg],
\end{eqnarray}
where $n_{ab}^{ij}\equiv n_{ab}^in_{ab}^j$ and two relations that
\begin{equation}
	\label{}
	\partial^2_{ij}\ln r_a = \frac{\delta_{ij}-2n_a^in_a^j}{r_a^2},
\end{equation}
and
\begin{equation}
	\label{}
	\partial_{ai}\partial_{bj}\ln S_{ab}=\frac{n_{ab}^{ij}-\delta^{ij}}{r_{ab}S_{ab}}+\frac{(n_{ab}^i-n_a^i)(n_{ab}^j+n_b^j)}{S_{ab}^2},
\end{equation}
are used.

In solving $Q_{ij}$, we can use the transformation (\ref{bgij2})-(\ref{bg002}) and set $\xi^i=c_{13}\Delta^{-1}(L_{i,t}-\overset{(3)}{K}_{i,t})$ to eliminate the term $c_{13}\Delta^{-1}(L_{i,jt}+L_{j,it}-\overset{(3)}{K}_{i,jt}-\overset{(3)}{K}_{j,it})$. After that, with the relations that
\begin{eqnarray}
	\label{}
	\Delta^{-1}\{-4\pi \mathcal{G} \sigma_{ij}\} & = & \sum_a \int \frac{\mathcal{G} m_a v_a^iv_a^j}{|\bm{x}-\bm{z}|}\delta(\bm{z}-\bm{y}_a)\mathrm{d}^3\bm{z}\nonumber\\
	& = & \sum_a \frac{\mathcal{G}m_a}{r_a}v_a^iv_a^j,
\end{eqnarray}
and
\begin{equation}
	\label{}
	\Delta^{-1}\{-4\pi \mathcal{G} \sigma_{kk}\}  =  \sum_a \frac{\mathcal{G}m_a}{r_a}v_a^2,
\end{equation}
transferred $Q^{(1)}_{ij}$ can be worked out as
\begin{eqnarray}
	\label{}
	Q^{(1)}_{ij} & = &+4\bigg(1-\frac{1}{2}c_{14}\bigg)\sum_a \frac{\mathcal{G}m_a}{r_a}v_a^iv_a^j +\bigg(1-\frac{1}{2}c_{14}\bigg)\sum_a\mathcal{G}^2m_a^2\frac{n_a^in_a^j}{r_a^2}\nonumber\\
	&&-4\bigg(1-\frac{1}{2}c_{14}\bigg)\sum_a\sum_{b\neq a} \mathcal{G}^2m_am_b\bigg[\frac{n_{ab}^{ij}}{r_{ab}S_{ab}}+\frac{(n_{ab}^i-n_a^i)(n_{ab}^j+n_b^j)}{S_{ab}^2}\bigg]\nonumber\\
	&&+\delta_{ij}\bigg\{+C^{Q_{ij}}_{N_{,tt}}\sum_a\frac{\mathcal{G}m_a}{r_a}\bigg[ v_a^2-(\bm{n}_a\cdot\bm{v}_a)^2\bigg]-4\bigg(1-\frac{1}{2}c_{14}\bigg)\sum_a\frac{\mathcal{G}m_av_a^2}{r_a}\nonumber\\
	&&\phantom{+\delta_{ij}\bigg\{}+\bigg(1+\frac{1}{2}c_{14}\bigg)\sum_a\frac{\mathcal{G}^2m_a^2}{r_a^2}+2\sum_a\sum_{b\neq a}\frac{\mathcal{G}^2m_am_b}{r_ar_b}\nonumber\\
	&&\phantom{+\delta_{ij}\bigg\{}+C^{Q_{ij}}_{N_{,tt}}\sum_a\sum_{b\neq a}\frac{\mathcal{G}^2m_am_b}{r_{ab}^2}\bm{n}_a\cdot\bm{n}_{ab}+4\bigg(1-\frac{1}{2}c_{14}\bigg)\sum_a\sum_{b\neq a} \frac{\mathcal{G}^2m_am_b}{r_{ab}S_{ab}}\bigg\}.
\end{eqnarray}

Collecting all these results together, we have the metric for 2PN light propagation as
\begin{eqnarray}
	\label{}
	g^{(1)}_{00} & = & -1+\epsilon^2\sum_a\frac{2\mathcal{G}m_a}{r_a}\nonumber\\
	&&+\epsilon^4\bigg\{+\sum_a\frac{\mathcal{G}m_a}{r_a}\bigg[(3+C^L_{N,_{tt}})v_a^2-C^L_{N,_{tt}}(\bm{n}_a\cdot\bm{v}_a)^2\bigg]-2\sum_a \frac{\mathcal{G}^2m_a^2}{r_a^2}\nonumber\\
		&&\phantom{+\epsilon^4\bigg\{}+\sum_a\sum_{b\neq a}\mathcal{G}^2m_am_b\bigg[-\frac{2}{r_ar_b}-\frac{2}{r_ar_{ab}}+C^{L}_{N_{,tt}}\frac{\bm{n}_a\cdot\bm{n}_{ab}}{r_{ab}^2}\bigg]\bigg\}\nonumber\\
	&&+\mathcal{O}(\epsilon^5),\\
	g^{(1)}_{0i} & = & +\epsilon^3\bigg\{-\frac{2c_{13}^2-2c_1c_{14}+8c_1}{c_3^2-c_1^2+2c_1}\sum_a\frac{\mathcal{G}m_av_a^i}{r_a}+\frac{2(c_3^2+c_1c_4)}{c_3^2-c_1^2+2c_1}\sum_a\frac{\mathcal{G}m_a}{r_a}(\bm{n}_a\cdot \bm{v}_a )n_a^i\bigg\}\nonumber\\
	&&+\mathcal{O}(\epsilon^5),\\
	g^{(1)}_{ij} & = &+\delta_{ij}+\epsilon^2\sum_a\frac{2\mathcal{G}m_a}{r_a}\delta_{ij}\nonumber\\
	&&+\epsilon^4\delta_{ij}\bigg\{\sum_a\frac{\mathcal{G}m_a}{r_a}\bigg[\bigg(C^{Q_{ij}}_{N_{,tt}}-1+2c_{14}\bigg)v_a^2-C^{Q_{ij}}_{N_{,tt}}(\bm{n}_a\cdot\bm{v}_a)^2\bigg]+\bigg(1+\frac{1}{2}c_{14}\bigg)\sum_a\frac{\mathcal{G}^2m_a^2}{r_a^2}\nonumber\\
	&&\phantom{+\epsilon^4\delta_{ij}\bigg\{}+\sum_a\sum_{b\neq a} \mathcal{G}^2m_am_b\bigg[\frac{2}{r_ar_b}-\frac{2}{r_ar_{ab}}+C^{Q_{,ij}}_{N_{,tt}}\frac{\bm{n}_a\cdot\bm{n}_{ab}}{r_{ab}^2}+\frac{2(2-c_{14})}{r_{ab}S_{ab}}\bigg]\bigg\}\nonumber\\
	&&+\epsilon^4\bigg\{+4\bigg(1-\frac{1}{2}c_{14}\bigg)\sum_a \frac{\mathcal{G}m_a}{r_a}v_a^iv_a^j +\bigg(1-\frac{1}{2}c_{14}\bigg)\sum_a\mathcal{G}^2m_a^2\frac{n_a^in_a^j}{r_a^2}\nonumber\\
	&&\phantom{+\epsilon^4\bigg\{}-4\bigg(1-\frac{1}{2}c_{14}\bigg)\sum_a\sum_{b\neq a} \mathcal{G}^2m_am_b\bigg[\frac{n_{ab}^{ij}}{r_{ab}S_{ab}}+\frac{(n_{ab}^i-n_a^i)(n_{ab}^j+n_b^j)}{S_{ab}^2}\bigg]\bigg\}\nonumber\\
	&&+\mathcal{O}(\epsilon^5).
\end{eqnarray}
When $c_1=c_2=c_3=c_4=0$, ae-theory goes back to GR, and the above metric reduces to the metric given in GR by \cite{Blanchet1998}. 

If we only consider the case that light just passes the limb of the Sun ($M_{\odot}$), which will provide the strongest light deflection effect in the solar system, we can simplify the 2PN metric further, by neglecting the contributions from planets. Hence, the metric for such a practical 2PN light deflection experiments can be simplified as
\begin{eqnarray}
	\label{}
	g^{(2)}_{00} & = & -1+\epsilon^2\frac{2\mathcal{G}M_{\odot}}{R_{\odot}}-2\epsilon^4\frac{\mathcal{G}^2M_{\odot}^2}{R_{\odot}^2},\\
	g^{(2)}_{0i} & = & 0,\\
	g^{(2)}_{ij} & = &+\delta_{ij}+\epsilon^2\frac{2\mathcal{G}M_{\odot}}{R_{\odot}}\delta_{ij}\nonumber\\
	&&+\epsilon^4\bigg[\delta_{ij}\bigg(1+\frac{1}{2}c_{14}\bigg)\frac{\mathcal{G}^2M_{\odot}^2}{R_{\odot}^2}+\bigg(1-\frac{1}{2}c_{14}\bigg)\mathcal{G}^2M_{\odot}^2\frac{n_{\odot}^in_{\odot}^j}{R_{\odot}^2}\bigg],
\end{eqnarray}
which is anisotropic in $g_{ij}$, causing unconvenience in the calculation of 2PN light deflection angle. Only one 2PN parameter $c_{14}$, which deviates from GR, remains in this special case.

\subsection{Isotropic coordinates for the dominated body}
The transformation between isotropic and harmonic coordinate system in GR involves only the radial coordinate, and is given by
\begin{equation}
	\label{}
	r_H+\epsilon^2\mathcal{G}m=r_I\bigg(1+\epsilon^2\frac{\mathcal{G}m}{2r_I}\bigg)^2,
\end{equation}
whose 2PN approximation is
\begin{equation}
	\label{}
	r_I=r_H-\epsilon^4\frac{\mathcal{G}^2m^2}{4r_H}+\mathcal{O}(\epsilon^6),
\end{equation}
where $r_I$ and $r_H$ represent the radial coordinate in the isotropic and harmonic system respectively. Hence, we obtain the transformation in ae-theory for spatial components as
\begin{equation}
	\label{}
	\bar{x}^i=x^i-\epsilon^4\bigg(1-\frac{1}{2}c_{14}\bigg)\frac{\mathcal{G}^2m^2}{4r}n^i,
\end{equation}
and, using the transformation (\ref{bgij2})-(\ref{bg002}) again, we get the metric $g^{(2)}_{\mu\nu}$ for 2PN light propagation as
\begin{eqnarray}
	\label{ds2ae}
	\mathrm{d}s^2 & = & \bigg(-1+\frac{2\mathcal{G}M_{\odot}}{c^2r}-\frac{2\mathcal{G}^2M_{\odot}^2}{c^4r^2}\bigg)c^2\mathrm{d}t^2\nonumber\\
	& &+\bigg[1+\frac{2\mathcal{G}M_{\odot}}{c^2r}+\bigg(\frac{3}{2}+\frac{1}{4}c_{14}\bigg)\frac{\mathcal{G}^2M_{\odot}^2}{c^4r^2}\bigg](\mathrm{d}^2r^2+r^2\mathrm{d}\Omega^2),
\end{eqnarray}
where $\mathrm{d}\Omega^2\equiv \mathrm{d}\theta^2+\sin^2\theta\mathrm{d}\phi^2$. In the spacetime with metric (\ref{ds2ae}), the light deflection angle up to 2PN approximation is
\begin{equation}
	\label{}
	\Delta \phi = \frac{4\mathcal{G}M_{\odot}}{c^2d}+\bigg[\frac{(15+c_{14})\pi}{4}-8\bigg]\frac{\mathcal{G}^2M_{\odot}^2}{c^4d^2},
\end{equation}
where $d$ represents the coordinate radius at the point of closest approach of the ray and $c_{14}$ is the \emph{only} non-GR parameter representing the deviation from GR, which poses another constraint on action parameters of ae-theory. 

Many previous works have been done to constrain the parameters $c_{1,2,3,4}$ by theoretical and experimental analyses, including the rate of primordial nucleosynthesis \cite{Carroll2004}, the rate of \^{C}erenkov radiation \cite{Elliott2005}, the stability and positive energy of the linearized wave modes \cite{Jacobson2004,Foster2006a}, the experimental bounds on the PPN parameters \cite{Eling2004,Foster2006a,Graesser2005}, which shows a two-parameter family of ae-theory can satisfy the requirements of observations (see \cite{Foster2006a} for a summary).  The constraints coming from the strong field effects and the rate of gravitational radiation damping in binary systems show that tests will be satisfied by the small-$c_{1,2,3,4}$ and the weak field PPN parameters conditions \cite{Foster2006b,Foster2006bE,Foster2007}.

In 1PN weak field experiments, the preferred frame PPN parameters are $|\alpha_1|\le 10^{-4}$ and $|\alpha_2|\le 10^{-7}$ \cite{TEGP,Will2006}. Putting both of them to zero leads to two conditions with two parameters \cite{Foster2006a}:
\begin{itemize}
	\item $c_2=(-2c_1^2-c_1c_3+c_3^2)/(3c_1)$, $c_4=-c_3^2/c_1$;
	\item $c_{13}=0$, $c_{14}=0$.
\end{itemize}
The latter case ($c_{14}=0$) will cause the propagation of the linearized wave modes of spin-0 and spin-1 with infinite velocities \cite{Jacobson2004,Foster2006a}. If it is shown that the difference between GR and ae-theory is extremely small by future 2PN light deflection experiments, it, together with the 1PN conditions given by \cite{Eling2004,Graesser2005,Foster2006a}, will improve the above conditions further ($\alpha_1=\alpha_2=0$ plus $c_{14}=0$):
\begin{itemize}
	\item $c_2=-2c_1/3$, $c_3=c_1$, $c_4=-c_1$;
	\item $c_{13}=0$, $c_{14}=0$,
\end{itemize}
which implies that the velocities of spin-0 and spin-1 modes waves are infinite in both cases if there is no deviation between ae-theory and GR in weak field experiments. Other experiments focusing on 2PN periastron advances in the binary pulsars and the solar system and on preferred frame effects can provide more constraints on the action parameters, $c_{1,2,3,4}$. These experiments need 2PN equations of motion in ae-theory. The simplest case is the free fall of a test particle in the Sun's gravitational field. It associated metric in the isotropic coordinate, which needs to extend $g_{00}$ in Eq. (\ref{ds2ae}) to $1/c^6$, is  
\begin{eqnarray}
	\label{ds2ae2}
	\mathrm{d}s^2 & = & \bigg[-1+\frac{2\mathcal{G}M_{\odot}}{c^2r}-\frac{2\mathcal{G}^2M_{\odot}^2}{c^4r^2}+\bigg(\frac{3}{2}-\frac{1}{12}c_{14}\bigg)\frac{\mathcal{G}^3M_{\odot}^3}{c^6r^3}\bigg]c^2\mathrm{d}t^2\nonumber\\
	& &+\bigg[1+\frac{2\mathcal{G}M_{\odot}}{c^2r}+\bigg(\frac{3}{2}+\frac{1}{4}c_{14}\bigg)\frac{\mathcal{G}^2M_{\odot}^2}{c^4r^2}\bigg](\mathrm{d}^2r^2+r^2\mathrm{d}\Omega^2),
\end{eqnarray}
where again only one non-GR parameter, $c_{14}$, appears in $g_{00}$ and $g_{ij}$.

\section{Conclusions}
In this paper, we obtain the 2PN approximation of ae-thoery by Chandrasekhar's approach. It shows a more comprehensive picture of the structure of 2PN approximation than scalar-tensor theories. Our works obtain five PPN parameters in 1PN, and they are consistent with previous works \cite{Eling2004,Graesser2005,Foster2006a}. Meanwhile, a 2PPN parameter in ae-theory, $\iota$, which is a non-GR parameter in the 2PN $g_{00}$, is discussed and compared with other theories, such as a multiscalar-tensor theory \cite{Damour1992,Damour1996} and a scalar-tensor theory with an intermediate range force \cite{Xie2007}. It is shown that $\iota$ is zero in ae-theory, which means that ae-theory, similar to GR, does not have the third nolinearity of Newtonian potential in $g_{00}$. For future applications in deep space laser ranging missions, we derive the 2PN metric for light propagation in the case of $N$ nonspinning point masses as a simplified model for the solar system.  The deviation of the resulting 2PN deflection angle of light between GR and ae-theory is dependent only on $c_{14}$, which gives another constraint of action parameters, $c_{1,2,3,4}$. If future experiments show $c_{14}$ is zero, it means that the linearized waves with the spin-0 and spin-1 modes in ae-theory will propagate with infinite velocities.

\begin{acknowledgements}
We acknowledge very useful and helpful comments and suggestions from our anonymous referee. We thank Xue-Mei Deng of Purple Mountain Observatory for her helpful discussions and advices. This work is funded by the Natural Science Foundation of China under Grants No. 10563001.
\end{acknowledgements}

\appendix

\section{Constants $C^{X}_{Y_{,\mu\nu}}$ and $\Delta \mathcal{A}$}
\label{cxy}

There are some constants like $C^{X}_{Y_{,\mu\nu}}$ in the coefficients of the metric, whose expressions are
\begin{eqnarray}
	\label{}
	C^{\overset{(3)}{K}_i}_{N_{,it}} & = & \frac{1}{c_3^2+2c_1-c_1^2}\bigg[c_4\bigg(\frac{3}{2}c_{13}-1\bigg)-\frac{c_{23}}{c_{123}}\bigg(c_{13}+c_3-c_4\bigg)\nonumber\\
	&&\phantom{\frac{1}{c_3^2+2c_1-c_1^2}\bigg[}+\frac{c_{13}}{c_{123}}\bigg(\frac{1}{2}c_{13}c_{23}+c_3^2-\frac{1}{2}c_2c_4-c_3c_4\bigg)\bigg],
\end{eqnarray}
\begin{eqnarray}
	\label{}
	C^L_{N_{,tt}} & = & \frac{1}{1-\frac{1}{2}c_{14}}\bigg[1-\frac{7}{2}c_1-\frac{9}{2}c_2-\frac{3}{2}c_3-2c_4\nonumber\\
	&&\phantom{\frac{1}{1-\frac{1}{2}c_{14}}\bigg[}+\frac{1}{2c_{123}}(c_2+c_{12}+c_{1234})^2\bigg],
\end{eqnarray}
\begin{equation}
	\label{}
	C^{Q_{ij}}_{N_{,tt}}  =  1-\frac{3}{2}c_{14}+c_{2}-\frac{(c_{12}+c_{24})(c_{23}-c_4)}{2c_{123}}+\frac{c_{14}}{2}C^L_{N_{,tt}},
\end{equation}
\begin{eqnarray}
	\label{}
	C^{Q_i}_{N_{,i}N_{,t}} & = & -\frac{1}{4c_{123}(c_3^2-c_1^2+2c_1)}\bigg[+6c_{2}c_{3}c_{14}+c_{3}c_{4}c_{234}+4c_{4}c_{3}^2+3c_{1}c_{4}^2\nonumber\\
	&&\phantom{-\frac{1}{4c_{123}(c_3^2-c_1^2+2c_1)}\bigg[}-c_{1}c_{4}c_{23}-2c_{1}c_{4}c_{12}-4c_{1}c_{4}c_{123}-2c_{3}^2c_{23}\nonumber\\
	&&\phantom{-\frac{1}{4c_{123}(c_3^2-c_1^2+2c_1)}\bigg[}-4c_{3}^2c_{123}-2c_{1}^2c_{14}-24c_{1}c_{123}\bigg]\\
	C^{Q_i}_{NN_{,it}} & = & \frac{1}{4c_{123}(c_3^2-c_1^2+2c_1)}\bigg[+c_{3}c_{4}^2+17c_{1}c_{4}c_{23}+18c_{1}^2c_{1234}-c_{1}c_{4}^2\nonumber\\
	&&\phantom{\frac{1}{4c_{123}(c_3^2-c_1^2+2c_1)}\bigg[}-2c_{3}^2c_{4}-c_{3}c_{4}c_{23}-2c_{3}^2c_{123}\bigg],
\end{eqnarray}
\begin{eqnarray}
	\label{}
	C^{\overset{(5)}{K}_i}_{N_{,i}N_{,t}} & = & -\frac{1}{8c_{123}^2(c_3^2-c_1^2+2c_1)}\bigg(+2c_{1}^4+14c_{1}^3c_{2}+8c_{1}^3c_{3}-14c_{1}^3c_{4}\nonumber\\
	&&+12c_{1}^2c_{2}^2+18c_{1}^2c_{2}c_{3}-13c_{1}^2c_{2}c_{4}-8c_{1}^2c_{3}^2-23c_{1}^2c_{3}c_{4}+5c_{1}^2c_{4}^2\nonumber\\
	&&-26c_{1}c_{2}c_{3}^2-8c_{1}c_{2}c_{3}c_{4}+4c_{1}c_{2}c_{4}^2-32c_{1}c_{3}^3-2c_{1}c_{3}^2c_{4}+8c_{1}c_{3}c_{4}^2\nonumber\\
	&&-12c_{2}^2c_{3}^2-30c_{2}c_{3}^3+5c_{2}c_{3}^2c_{4}+4c_{2}c_{3}c_{4}^2-18c_{3}^4+7c_{3}^3c_{4}+3c_{3}^2c_{4}^2\nonumber\\
	&&-32c_{1}^3-80c_{1}^2c_{2}-96c_{1}^2c_{3}+12c_{1}^2c_{4}-48c_{1}c_{2}^2-124c_{1}c_{2}c_{3}-4c_{1}c_{2}c_{4}\nonumber\\
	&&-76c_{1}c_{3}^2+4c_{1}c_{3}c_{4}-4c_{1}c_{4}^2-12c_{2}^2c_{3}-14c_{2}^2c_{4}-24c_{2}c_{3}^2-24c_{2}c_{3}c_{4}\nonumber\\
	&&-2c_{2}c_{4}^2-12c_{3}^3-10c_{3}^2c_{4}-2c_{3}c_{4}^2\bigg),
\end{eqnarray}
\begin{eqnarray}
	\label{}
	C^{\overset{(5)}{K}_i}_{NN_{,it}} & = & \frac{1}{8c_{123}^2(c_3^2-c_1^2+2c_1)}\bigg(+18c_{1}^4+28c_{1}^3c_{2}+46c_{1}^3c_{3}+32c_{1}^3c_{4}+10c_{1}^2c_{2}^2\nonumber\\
	&&+58c_{1}^2c_{2}c_{3}+51c_{1}^2c_{2}c_{4}+50c_{1}^2c_{3}^2+73c_{1}^2c_{3}c_{4}-c_{1}^2c_{4}^2+20c_{1}c_{2}^2c_{3}\nonumber\\
	&&+20c_{1}c_{2}^2c_{4}+52c_{1}c_{2}c_{3}^2+80c_{1}c_{2}c_{3}c_{4}+34c_{1}c_{3}^3+48c_{1}c_{3}^2c_{4}+10c_{2}^2c_{3}^2\nonumber\\
	&&+20c_{2}^2c_{3}c_{4}+22c_{2}c_{3}^3+29c_{2}c_{3}^2c_{4}+12c_{3}^4+7c_{3}^3c_{4}+c_{3}^2c_{4}^2+16c_{1}^2c_{2}\nonumber\\
	&&+16c_{1}^2c_{3}-20c_{1}^2c_{4}+16c_{1}c_{2}^2+28c_{1}c_{2}c_{3}-16c_{1}c_{2}c_{4}+12c_{1}c_{3}^2-16c_{1}c_{3}c_{4}\nonumber\\
	&&-4c_{2}^2c_{3}+2c_{2}^2c_{4}-8c_{2}c_{3}^2+8c_{2}c_{3}c_{4}-2c_{2}c_{4}^2-4c_{3}^3+6c_{3}^2c_{4}-2c_{3}c_{4}^2\bigg),
\end{eqnarray}
\begin{eqnarray}
	\label{}
	C^{\overset{(5)}{K}_i}_{L_{,it}} & = & \frac{1}{2c_{123}(c_3^2-c_1^2+2c_1)}\bigg(+2c_{1}^2c_{134}+c_{1}^2c_{234}+c_{3}^2c_{234}\nonumber\\
	&&+2c_{1}c_{23}c_{34}+2c_{12}c_{3}c_{4}+2c_{1}c_{23}-2c_{1}c_{4}\bigg),
\end{eqnarray}
\begin{eqnarray}
	\label{}
	C^Q_{NN_{,tt}} & = & \frac{2}{2-c_{14}}\bigg[2-\frac{1}{4}c_1-c_2+\frac{1}{4}c_3-\frac{1}{8}c_4\nonumber\\
	&&+\frac{(c_{12}+c_{24})(7c_1+2c_2+3c_3+5c_4)}{4c_{123}}-\frac{c_4(c_{12}+c_{24})^2}{8c_{123}^2}\bigg],
\end{eqnarray}
\begin{eqnarray}
	\label{}
	C^Q_{N_{,t}N_{,t}} & = & \frac{2}{2-c_{14}}\bigg[4-2c_1-\frac{5}{2}c_2+\frac{1}{4}c_3-\frac{15}{8}c_4\nonumber\\
	&&+\frac{(c_{12}+c_{24})(10c_1+6c_3+5c_{24})}{4c_{123}}-\frac{c_4(c_{12}+c_{24})^2}{8c_{123}^2}\bigg],
\end{eqnarray}
\begin{eqnarray}
	\label{}
	C^Q_{N_{,kt}L_k} & = & -\frac{2}{2-c_{14}}\bigg\{2-2c_1-\frac{1}{2}c_{24}-\frac{3}{2}c_3\nonumber\\
	&&-\frac{2}{c_3^2-c_1^2+2c_1}\bigg[(c_1+2c_3-c_4)(c_3^2+c_1c_4)-c_4c_{13}\bigg(\frac{3}{2}c_{13}-1\bigg)\bigg]\nonumber\\
	&&+\frac{2c_{13}^2}{c_{123}(c_3^2-c_1^2+2c_1)}\bigg(\frac{1}{2}c_{13}c_{23}+c_3^2-\frac{1}{2}c_2c_4-c_3c_4\bigg)\nonumber\\
	&&-\frac{2c_{13}c_{23}(c_{13}+c_3-c_4)}{c_{123}(c_3^2-c_1^2+2c_1)}\bigg\},
\end{eqnarray}
and
\begin{eqnarray}
	\label{}
	C^Q_{N_{,kt}\overset{(3)}{K}_k} & = &-\frac{2}{2-c_{14}}\bigg\{ c_{14}-c_2+\frac{(c_{12}+c_{24})(c_{23}-c_4)}{2c_{123}}\nonumber\\
	&&+\frac{2}{c_3^2-c_1^2+2c_1}\bigg[(c_1+2c_3-c_4)(c_3^2+c_1c_4)-c_4c_{13}\bigg(\frac{3}{2}c_{13}-1\bigg)\bigg]\nonumber\\
	&&+\frac{c_{13}^2(c_2c_4+2c_3c_4-c_{13}c_{23}-2c_3^2)}{c_{123}(c_3^2-c_1^2+2c_1)}+\frac{2c_{13}c_{23}(c_{13}+c_3-c_4)}{c_{123}(c_3^2-c_1^2+2c_1)}\bigg\}.
\end{eqnarray}

In the Eq. (\ref{withA}), we introduce a variable $\mathcal{A}$, which satisfies a Poisson's equation as
\begin{eqnarray}
	\label{}
	\Delta\mathcal{A} & = & \bigg[-\frac{3}{2}+\frac{1}{8c_{123}}(10c_{13}+12c_2-3c_4)+\frac{c_4(c_{12}+c_{24})}{8c_{123}^2}\bigg]N\Delta N_{,t}\nonumber\\
	&&+\bigg[\frac{2c_3-5c_4}{8c_{123}}+\frac{c_{12}+c_{24}}{2c_{123}^2}\bigg(\frac{1}{2}c_1+c_3-\frac{1}{4}c_4\bigg)\bigg]N_{,t}\Delta N\nonumber\\
	&&-\frac{c_{13}}{4c_{123}}N_{,k}\Delta L_k+\frac{c_1+2c_3-c_4}{2c_{123}}\bigg(L_k\Delta N_{,k}-\overset{(3)}{K}_k\Delta N_{,k}\bigg)\nonumber\\
	&&-\frac{c_4}{2c_{123}}N_{,k}\Delta\overset{(3)}{K}_k+\frac{c_{13}}{c_{123}}\bigg(\frac{1}{2}+\frac{1}{4}c_{14}\bigg)N_{,k}N_{,kt}+\frac{4c_{13}(2-c_{14})}{c_{123}}\pi \mathcal{G} \sigma_{kk,t}\nonumber\\
	&&-\frac{4c_{13}}{c_{123}}\pi \mathcal{G} (N\sigma_{,t}+\sigma N_{,t})-\frac{c_{14}}{c_{123}}\bigg(\frac{3}{2}-\frac{c_{12}+c_{24}}{2c_{123}}\bigg)N_{,ttt}\nonumber\\
	&&+\frac{c_{13}}{2c_{123}}N_{,ttt}\bigg[\frac{7}{2}-\frac{9}{4}c_{14}+\frac{3}{2}c_2+\frac{1}{2}c_{13}+\frac{c_{12}+c_{24}}{4c_{123}}(2c_1-3c_2-c_3+3c_4)\nonumber\\
	&&\phantom{+\frac{c_{13}}{2c_{123}}N_{,ttt}\bigg[}+\frac{2}{2-c_{14}}\bigg(\frac{1}{2}-\frac{3}{4}c_{14}\bigg)\bigg(-1+\frac{7}{2}c_1+\frac{9}{2}c_2+\frac{3}{2}c_3+2c_{4}\bigg)\nonumber\\
	&&\phantom{+\frac{c_{13}}{2c_{123}}N_{,ttt}\bigg[}+\frac{2}{c_{123}(2-c_{14})}\bigg(\frac{3}{8}c_{14}-\frac{1}{4}\bigg)(c_2+c_{12}+c_{1234})^2\bigg].
\end{eqnarray}

\newpage %Just because of unusual number of tables stacked at end
\bibliography{all}% Produces the bibliography via BibTeX.

\end{document}